# A model of heat transfer from a cylinder in high-speed slip flow and determination of temperature jump coefficients using hot-wires

Diogo C. Barros, Lionel Larchevêque, Pierre Dupont

*Aix Marseille Université, CNRS, IUSTI, Marseille, France*

**Abstract**

In small-scale, low-density or high-speed flows, the mean free path of the gas and its molecular interaction with a solid interface are key properties for the analysis of heat transfer mechanisms occurring in many flow processes ranging from micro-scale to aerospace applications. Here, we specifically examine the effects of temperature jump at the gas-solid interface on the convection from a cylinder in the high-speed slip flow regime. By employing the classical Smoluchowski temperature jump condition, we derive a simple model that explicitly predicts the heat flux (Nusselt number Nu) from the surface of a small heated cylinder simulating a hot-wire as a function of the Knudsen number (Kn) and the thermal (or energy) accommodation coefficient ($\sigma_T$) of the gas molecules interacting with the surface. The model, derived analytically and verified empirically by numerical simulations, helps clarifying coupled gas rarefaction and temperature effects on the heat transfer from a cylinder in high-speed flow. In addition, we employ the model reversely to propose a novel methodology to compute gas-surface thermal accommodation or temperature jump coefficients from hot-wire measurements.

## 1. Introduction

Forced convection over moving objects in rarefied gases is relevant for many applications including aerodynamic heat transfer in high-altitude flight, atmosphere spacecraft reentry or Mars landing, gas-particle dynamics in compressible flows and hot-wire measurements in low-density environment (Comte-Bellot, 1976; Xie et al., 2017, 2018; Roseman and Argrow, 2021; Pointer et al., 2023; Capecelatro and Wagner, 2024). In those conditions, the gas mean free path and the gas-surface thermal exchange effectiveness play a central role on macroscopic heat transfer mechanisms.

The effects of gas rarefaction becomes significant when the molecular mean free path $\lambda$ is comparable to a relevant flow length scale $L$. The mean free path of the gas can be written as $\mu = c\rho \bar{v} \lambda$ for a hard elastic sphere model, where $2c \approx 1$, $\mu$ is the gas dynamic viscosity, and $\bar{v}$ is the mean molecular speed $\bar{v} = \sqrt{8RT/\pi}$ at temperature $T$ for a defined gas constant $R$ (Kennard, 1938). To quantify the degree of rarefaction, the Knudsen number is defined as $\text{Kn} = \lambda/L$, with the continuum fluid assumption valid for $\text{Kn} \to 0$. The Knudsen number is hence related to more familiar flow parameters by (Karniadakis et al., 2006)

$$\text{Kn} = \sqrt{\frac{\gamma\pi}{2}} \frac{\text{Ma}}{\text{Re}}. \qquad (1)$$

In this relation, $\text{Re} = \rho u L/\mu$ is the Reynolds number defined using the fluid density $\rho$, and the flow velocity $u$. The Mach number $\text{Ma} = u/a$ compares the flow velocity with the sound speed $a = \sqrt{\gamma RT}$. The heat capacity ratio is denoted by $\gamma$.

In slightly rarefied conditions ($0.01 < \text{Kn} < 0.1$), the fluid slips over a surface, and its dynamics can be modelled by the Navier-Stokes equations subject to velocity slip and temperature jump boundary conditions. This is called the slip flow regime (Kennard, 1938; Akhlaghi et al., 2023). In seminal work, James C. Maxwell provided the first model accounting for a slip velocity nearby the surface (Maxwell, 1879). In his model, the fluid velocity adjacent to the wall is proportional to the mean free path of the gas, approaching no slip in the continuum regime. Two decades later, in analogy to Maxwell's theory, von Smoluchowski proposed a temperature jump across the heated wall equally proportional to $\lambda$, but a decreasing function of the thermal (or energy) accommodation coefficient $\sigma_T$ that determines the effectiveness of gas-surface energy exchange (Smoluchowski, 1898; Chambre and Schaaf, 2017).

**Nomenclature**

*Symbols*
- a    Speed of sound
- A    King's law coefficient
- B    King's law coefficient
- d    Wire diameter
- e    Internal energy
- H    Averaged heat loss per unit length
- h    Convective heat transfer coefficient
- k    Gas thermal conductivity
- Kn    Knudsen number
- l    Wire length
- L    Relevant flow length scale
- Ma    Mach number
- n    King's law exponent
- P    Gas pressure
- Pr    Prandtl number
- q    Heat flux
- R    Gas constant
- Re    Reynolds number
- T    Gas temperature
- u    Local streamwise flow velocity
- v    Local transverse flow velocity
- $\bar{v}$    Mean molecular speed

*Greek symbols*
- $\sigma_M$    Momentum accommodation coefficient
- $\sigma_T$    Thermal accommodation coefficient
- $\lambda$    Mean free path
- $\lambda_e$    Equivalent free path
- $\delta_s$    Bow shock distance
- $\Delta$    Slip length
- $\varepsilon$    Empirical model parameter
- $\phi$    Correction prefactor
- $\gamma$    Gas heat capacity ratio
- $\mu$    Fluid dynamic viscosity
- $\rho$    Fluid density
- $\tau$    Wire overheat ratio
- $\tau_{i,j}$    Fluid stresses
- $\zeta_T$    Temperature jump coefficient

*Subscripts*
- c    Continuum condition
- g    Gas or near-wall condition
- r    Recovery or reference condition
- w    Wire or wall condition
- o    Stagnation condition
- $\infty$    Free-stream condition
- $*$    Post-shock condition

It is hence fundamental to evaluate the impact of both $\lambda$ and $\sigma_T$ on the convective heat transfer coefficient from a given surface.

We consider here the heat flux from a small cylinder simulating a hot-wire with diameter $d$ exposed to known flow conditions. Specifically for this flow, the relevant length scale is the wire diameter ($L = d$) such that $\mathrm{Kn} = \lambda/d$. The Nusselt number, or dimensionless heat flux, is defined as $\mathrm{Nu} = hd/k$, where $h$ is the average convective heat transfer coefficient, and $k$ the surrounding gas thermal conductivity. For a circular cylinder, the Nusselt number assumes the following form:

$$\mathrm{Nu} = \frac{H}{\pi k (T_w - T_r)}, \qquad (2)$$

where $H$ is the averaged heat loss per unit length, and $T_w$ is the isothermal wall temperature. $T_r$ is the reference temperature, usually the recovery or equilibrium temperature of the cylinder for a given flow condition in the absence of heat transfer (Bruun, 1995).

From hot-wire measurements in supersonic flow, Kovasznay (1950) and Laufer and McClellan (1956) demonstrated that the normalized heat transfer follows the empirical law $\mathrm{Nu} = A + B\mathrm{Re}^n$ in the Reynolds number range $20 < \mathrm{Re} < 300$ for Mach numbers greater than 1.2. The coefficients $A$ and $B$ are decreasing functions of the wire overheat ratio $\tau = (T_w - T_r)/T_o$ defined using the stagnation temperature $T_o$. In particular, their results present a compelling $n = 0.5$ exponent, in close agreement with the seminal King's law for incompressible flows (King, 1914). Later compilations of convective heat transfer reported by Baldwin et al. (1960) and Dewey Jr (1965) indicated a smooth transition from $n = 0.5 \to 1$ for supersonic flows at $\mathrm{Re} = \mathcal{O}(1)$. While $\mathrm{Nu} \propto \mathrm{Re}^{0.5}$ corresponds to a continuum, laminar boundary layer heat transfer scaling from the wire surface to the main flow, the unitary exponent is attributed to free-molecule rarefied gas effects (Stalder et al., 1951, 1952; Dewey Jr, 1961; Weltmann, 1960).

Although acknowledged in previous investigations (Collis and Williams, 1959; Baldwin et al., 1960; Andrews et al., 1972), it is crucial to quantify the impact of both Kn and $\sigma_T$ on the heat transfer in the slip flow regime, in particular for high-speed flows where those correlations have not been neither reported nor demonstrated comprehensively. Here, we employ direct numerical simulations and conduct wind-tunnel experiments in supersonic flow to understand the sensitivity of the heat flux to the Knudsen number and the gas-surface thermal accommodation coefficient. A heat transfer model is derived based on the numerical simulations, and employed to obtain both the accommo-dation and the temperature jump coefficients from hot-wire measurements. The numerical and experimental methods are described in section 2. The heat transfer results and the derivation of a heat transfer model are discussed in section 3, where an immediate outcome of the model is a novel methodology that we propose to determine the gas-surface thermal accommodation coefficient from classical hot-wire measurements. The concluding remarks are presented in section 4.

## 2. Methods

### 2.1. Wind-tunnel experiments

The experiments were performed in the supersonic wind-tunnel of the Institut Universitaire des Systèmes Thermiques et Industriels, Marseille, France. This facility can be operated for several hours with negligible changes of stagnation conditions. The plenum pressure was monitored and controlled in the range $0.2 \leq P_o \leq 0.8$ bar, while the established total temperature in the wind-tunnel chamber was equal to the room temperature $T_o = 298$ K with variations smaller than 0.4%. A converging-diverging nozzle generated the free-stream flow with designed Mach number $Ma_\infty = 2$ in the 170 mm × 105 mm test section.

Heat loss measurements from a fine wire were conducted using a Dantec Streamline constant-temperature hot-wire anemometer system with a bridge configuration 1:1 and top bridge resistance 20 Ω. The probe was a platinum-plated tungsten cylindrical wire with nominal diameter $d = 5$ μm and length $l = 800$ μm. The tests consisted of systematic variations of both the Reynolds and the Mach numbers for a fixed wire temperature $T_w = 495 \pm 3$ K. The wire was maintained at a constant temperature $T_w$ by adjusting the bottom electrical resistance in the anemometer circuit. The anemometer hence balances the Wheatstone bridge such that the wire resistance matches the adjusted bottom resistance. The value of the wire resistance was $4.47 \pm 0.02$ Ω for all tests. From this value of resistance, we calculated the wire temperature $T_w$ using the material temperature coefficients of resistivity (Bruun, 1995). The Reynolds number was varied by modifying the stagnation pressure of the wind-tunnel, while the Mach number was changed by traversing the hot-wire in the streamwise direction along the diverging section of the nozzle. The resulting aerodynamic conditions covered in the experiments are $15 \leq Re_\infty \leq 80$ and $1.3 \leq Ma_\infty \leq 2$. For each condition, we measured the equilibrium temperature of the wire $T_r$ to compute the overheat ratio. In our measurements, $\tau = 0.7$ within 2% due to slight variations of $T_r$ and $T_o$.

Finally, the Nusselt number was computed using equation 2 by evaluating the fluid properties at post-shock conditions, accounting for the detached shock wave upstream of the cylindrical wire (Kovasznay, 1950). End-conduction effects were considered using the correction method described in Kovasznay and Tormarck (1950). The resulting uncertainty on the value of the Nusselt number is ±0.05.

### 2.2. Direct numerical simulations

To quantify the effects of gas rarefaction on the heat flux from the cylinder, direct numerical simulations of the flow past a cylindrical wire were performed to compare the Nusselt number with both no-slip and temperature jump boundary conditions. The simulations were performed with a free-stream Mach number $Ma_\infty = 2$, $50 \leq Re_\infty \leq 400$ and variable overheat ratio $\tau$. Given the fact that conventional hot-wire measurements are usually corrected for end-conduction effects, the resulting heat transfer data is nominally two-dimensional, as previously considered in past simulations of thermal anemometry data (Shi et al., 2004; Ikeya et al., 2017). The supersonic flow around the small cylinder was modeled by the continuity, the full two-dimensional compressible Navier-Stokes and energy equations as follows:

$$\frac{\partial \rho}{\partial t} + \frac{\partial (\rho u)}{\partial x} + \frac{\partial (\rho v)}{\partial y} = 0, \quad (3)$$

$$\frac{\partial (\rho u_j)}{\partial t} + \frac{\partial (\rho u_j u_i)}{\partial x_i} = -\frac{\partial p}{\partial x_i} + \frac{\partial \tau_{ij}}{\partial x_i}, \quad (4)$$

$$\frac{\partial (\rho e_t)}{\partial t} + \frac{\partial (\rho u_i e_t)}{\partial x_i} = -\frac{\partial (u_i p)}{\partial x_i} + \frac{\partial (\tau_{ij} u_j)}{\partial x_i} - \frac{\partial q_i}{\partial x_i}, \quad (5)$$

$$\tau_{ij} = \mu \left( \frac{\partial u_i}{\partial x_j} + \frac{\partial u_j}{\partial x_i} \right) - \delta_{ij} \frac{2}{3} \mu \frac{\partial u_k}{\partial x_k}. \quad (6)$$

Here, $x_i$ are the space coordinates $(x, y)$, $t$ is the time and $u_i$ are the velocity components $(u, v)$. The flow pressure and stresses are denoted by $p$ and $\tau_{ij}$, respectively. Finally, $e_t = e + \rho \Sigma u_i^2 / 2$, where $e$ is the internal energy and $q_i$ the heat flux in the $i$ direction. With respect to the boundary conditions, we employed the Maxwell and the von Smoluchowski jump conditions:

$$u_g = \lambda \frac{2 - \sigma_M}{\sigma_M} \frac{\partial u}{\partial n}\bigg|_{wall} \quad (7)$$

and

$$T_g - T_w = \Delta \frac{\partial T}{\partial n}\bigg|_{wall}. \quad (8)$$

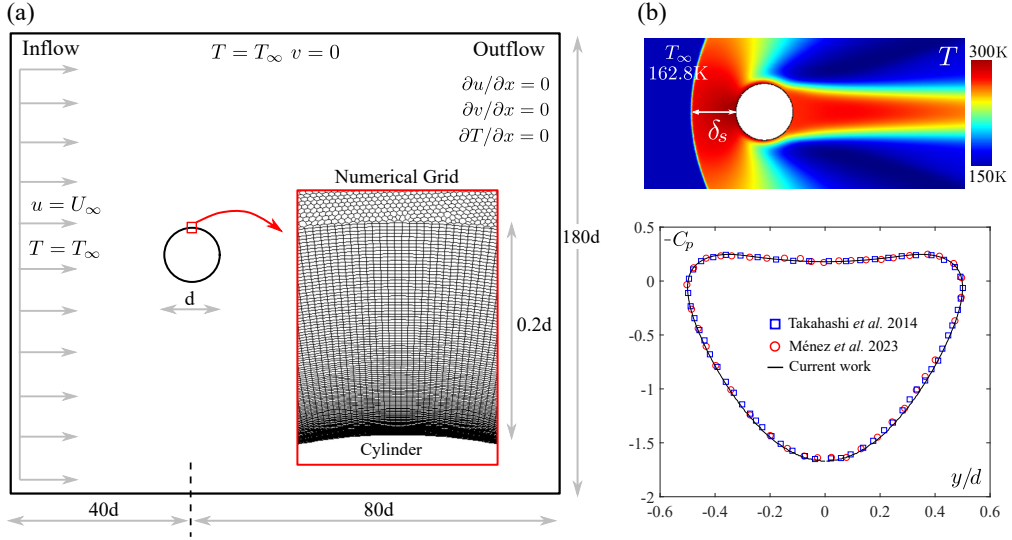

Figure 1: (a) Numerical domain and boundary conditions around the cylinder. (b) Results from the validation test case against the numerical simulations reported in Takahashi et al. (2014) and Ménez et al. (2023).

In the equations above, $u_g$ is the gas velocity adjacent to the wall, $\partial u/\partial n$ is the velocity gradient normal to the wall, and $\sigma_M$ is the tangential momentum accommodation coefficient of the Maxwell slip velocity condition set to $\sigma_M = 1$, assuming full accommodation. Due to the fact that the cylinder wall is isothermal, the thermal creep term vanishes for the present simulations (Martin and Boyd, 2006). Concerning the temperature jump, $T_g$ is the gas the temperature at the vicinity of the cylinder's surface, and $\Delta$ is the slip length that is a function of the thermal accommodation $\sigma_T$ and the Prandtl number Pr (Kennard, 1938):

$$\Delta = \frac{2-\sigma_T}{\sigma_T} \frac{2\gamma}{\gamma+1} \frac{\lambda}{\Pr}. \quad (9)$$

The compressible flow equations were solved in cartesian coordinates using the finite–volume solver StarCCM+ with a 2D unstructured mesh. For optimal accuracy, a pseudo–structured mesh aligned with the wall geometry was built from an even circumferential distribution of 720 wall segments, from which 120 layers of trapezoidal cells are extruded in the wall-normal direction with geometrically increasing thicknesses, for a total height of one-fifth of the cylinder's diameter. This thickness is large enough to cover the full extent of the thermal boundary layer at high heat flux locations over the cylinder. For all computations described hereafter with the no-slip boundary condition, cells adjacent to the wall have been found to have dimensions in wall units lower than 0.5 and 0.05 in the tangential and wall–normal directions, respectively. The mesh was also refined in the wake and in the region upstream of the cylinder where a bow shock exhibits a significant curvature. In figure 1(a), we present the computational domain around the circular cylinder representing the hot-wire with diameter $d$ and the boundary conditions of the domain. The inset shows the grid points set to simulate the boundary layer over the cylinder's surface over a distance of one-fifth of the cylinder diameter.

The numerical scheme relies on a third–order upwind–biased MUSCL reconstruction proposed by van Leer (1979), limited using the WENO strategy (Liu et al., 1994), while the convective fluxes are computed using the Roe scheme (Weiss and Smith, 1995). For all overheat ratios, the flow configuration having the largest Reynolds number, being therefore more prone to develop an unsteady von Kármán vortex street, was first solved by considering the unsteady Navier–Stokes equations 3–5. They were integrated over time in an implicit way with a second–order accuracy for a constant time step resulting in Courant number ranging from 50 to 100, up to reach a steady state. The flow configuration was then recomputed considering the steady Navier–Stokes equations. It was checked that both approaches resulted in similar flow fields, with local heat fluxes differing by less than 1‰. All the remaining flow configurations at lower Reynolds numbers were hence computed from the steady Navier–Stokes equations. For all computations, the wall temperature was set to a constant value $T_w$, and $T_r$ was determined by iteration to achieve an average null heat flux ($H = 0$).

We further checked grid resolution convergence for the most critical flow configurations, i.e. the largest Reynolds number and overheat ratio. It relies on a mesh strongly refined in the near-wall and upstream regions by a factor ranging from 2 to 25. It enables the shock structure to be solved from the mesh even for the largest Reynolds numbers. The values of convective heat fluxes varied no more than 2‰ in more refined meshes. Similar very low variations were found when substituting the AUSM$^+$ scheme (Liou, 1996) for the Roe one to compute the convective fluxes. Figure 1(b) depicts numerical validation results from a simulation with $\mathrm{Ma}_\infty = 2$, $\mathrm{Re}_\infty = 300$ and adiabatic boundary condition over the cylinder. The temperature field shown in the colormap highlights the location of the detached bow shock, at a distance $\delta_s \sim 0.77d$ upstream of the cylinder. This value is in good agreement with the simulations from Ménez et al. (2023), where they reported $\delta_s \sim 0.73d$. In addition, we present the pressure coefficient $C_p = (p - p_\infty)/(0.5\rho_\infty U_\infty^2)$ over the cylinder, and excellent agreement was obtained when compared to the pressure profiles from Takahashi et al. (2014) and Ménez et al. (2023).

## 3. Results and discussion

### 3.1. Effects of $\sigma_T$ and $\tau$ on the Nusselt number

In figure 2, we present the experimental and the numerical values of the Nusselt number alongside the heat transfer measurements from Laufer and McClellan (1956) at $\tau = 0.7$. Our experimental measurements agree very closely with the heat loss data from Laufer and McClellan (1956), and both experimental data sets lie in between the numerical simulations accounting for the temperature jump with thermal accommodation coefficients $\sigma_T = 1$ and $\sigma_T = 0.6$. A rough extrapolation of the three simulation curves towards larger Reynolds numbers indicates the absence of slip effects virtually at $\mathrm{Re} \sim 10^3$, i.e. $\mathrm{Kn} \sim \mathscr{O}(0.001)$, corresponding well to the continuum flow regime. When applying the temperature jump condition over the cylinder's surface, the heat loss decreases significantly with decreasing $\sigma_T$.

The effects of the temperature loading are presented in figure 3, where the no-slip, continuum Nusselt number $\mathrm{Nu}_c$ is plotted for multiple values of $\tau$ at selected Reynolds numbers. The continuum heat loss increases with $\tau$ following the relation $\mathrm{Nu}_c = A_c(\tau) + B_c(\tau)\sqrt{\mathrm{Re}}$, with $A_c(\tau)$ and $B_c(\tau)$ increasing functions of $\tau$ obtained from linear interpolations of $\mathrm{Nu}_c$. It is also shown in figure 3 the coupled effect of $\sigma_T$ and $\tau$ on Nu. The increase of temperature jump, by decreasing $\sigma_T$, has an opposite effect when compared to the temperature loading. This is notably remarked by the fact that Nu is almost independent of $\tau$ for $\sigma_T = 0.6$. Similar behavior was noted for other Reynolds numbers not shown here for brevity.

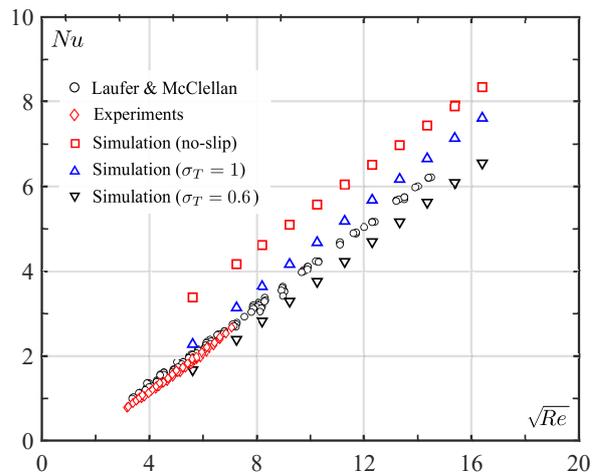

Figure 2: Experimental and numerical heat loss alongside the measurements from Laufer and McClellan (1956). The Nusselt and the Reynolds numbers were evaluated with the fluid properties at post-shock conditions for a fixed overheat ratio $\tau = 0.7$.

Our results highlight the crucial impact of $\sigma_T$ on the heat loss from the wire. Although previously acknowledged in low-speed flows (Collis and Williams, 1959; Andrews et al., 1972), a direct link between Nu, $\sigma_T$ and Kn has not been quantitatively described, particularly in high-speed flow conditions. It is perhaps of interest to remark that the thermal accommodation $\sigma_T$ critically depends on the gas-surface interface and wall temperature, and much disagreement is found in the literature about the correct values of $\sigma_T$ for a given gas-solid interface (Andrews et al., 1972).

### 3.2. Heat transfer model

To gain further insight on the impact of $\sigma_T$ on Nu, we follow Collis and Williams (1959) to write the convective heat flux from the wire using either its temperature $T_w$ or the gas temperature $T_g$ at the vicinity of wall:

$$q = h_w(T_w - T_r) = h_g(T_g - T_r). \qquad (10)$$

The local convective heat transfer coefficients $h_w$ and $h_g$ correspond to the slip flow and the continuum, no-slip flow models, respectively. By the definition of heat flux we write

$$q = -k\frac{\partial T}{\partial n}\Big|_{T_g}, \qquad (11)$$

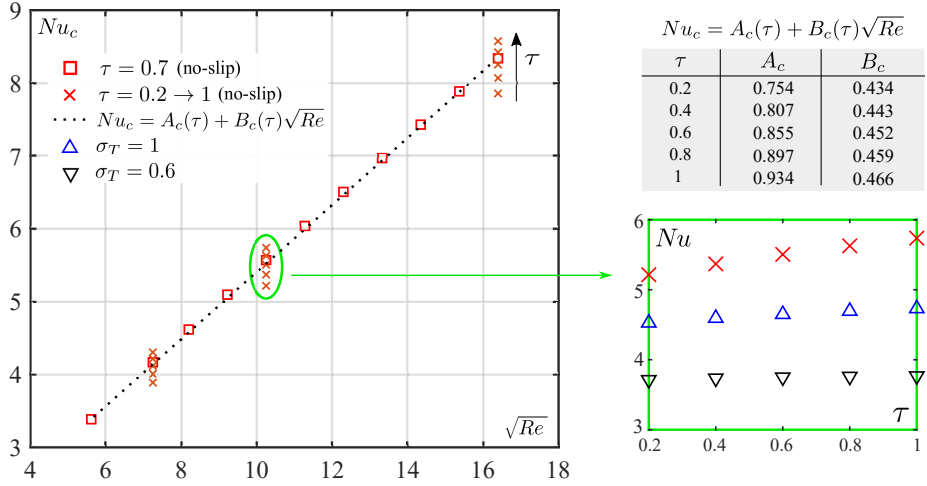

Figure 3: Effect of the overheat ratio $\tau$ on the Nusselt number under no-slip and slip flow conditions. The continuum Nusselt number is denoted by $Nu_c$, while Nu stands for Nusselt numbers obtained using the temperature jump condition with $\sigma_T = 1$ and $\sigma_T = 0.6$. The continuum heat loss increases with $\tau$ following $Nu_c = A_c(\tau) + B_c(\tau)\sqrt{Re}$. The inset shows the evolution of Nu against $\tau$ for Re $\sim$ 105 indicated by the circled data.

and, from the temperature jump condition of equation 8 eliminating the temperature gradient, we obtain:

$$\frac{h_g \Delta_{T_g}}{k_{T_g}} = \frac{T_w - T_g}{T_g - T_r}. \quad (12)$$

Replacing the slip length defined in equation 9, it follows that

$$h_w = \frac{h_g}{1 + \theta \frac{\lambda_{T_g}}{k_{T_g}} h_g}, \quad (13)$$

where $\theta = \Delta_{T_g}/\lambda_{T_g}$.

Equation 13 indicates a decrease of the heat transfer coefficient when compared to the no-slip condition, i.e. $h_w < h_g$. Given the impossibility of evaluating $\lambda_{T_g}$ at the vicinity of the entire wire's surface (Sharipov and Seleznev, 1998), it is tempting to suggest, after multiplication of equation 13 by $d/k$, that the average Nusselt number takes the empirical form

$$\mathrm{Nu} = \frac{\mathrm{Nu}_c}{1 + \phi \mathrm{Kn}_\infty \mathrm{Nu}_c}, \quad (14)$$

where the prefactor

$$\phi = \frac{1}{\mathrm{Pr}} \frac{2 - \sigma_T}{\sigma_T} \frac{2\gamma}{\gamma + 1} \left(\frac{T_w}{T^*}\right)^\varepsilon \quad (15)$$

assumes $T_g \sim T_w$, and accounts for the evaluation of the Nusselt number with the fluid properties at the post-shock temperature $T^*$, the ratio of them raised to the empirical parameter $\varepsilon$ to be determined from the numerical data. The free-stream Knudsen number $\mathrm{Kn}_\infty \propto$ $\mathrm{Ma}_\infty/\mathrm{Re}_\infty$ is employed considering $\lambda_\infty$ as the appropriate length scale when compared to the cylinder's diameter (Gatski and Bonnet, 2013).

The scaling model of equation 14 suggests a correction from the continuum heat transfer coefficient to the Nusselt number in the slip flow regime. This type of model has been proposed in many previous investigations of heat transfer over spheres and cylinders (Kavanau, 1955; Collis and Williams, 1959; Andrews et al., 1972), but its detailed verification accounting for systematic changes of the thermal accommodation coefficient could not have been reported given the difficulty of providing simultaneous values of $\mathrm{Nu}_c$, Nu and $\sigma_T$ for a single experiment.

To verify the validity of the model, equations 14 and 15 are tested with the results issued from the numerical simulations presented in figures 2 and 3. The continuum Nusselt number was approximated by the previous correlation $\mathrm{Nu}_c = A_c(\tau) + B_c(\tau)\sqrt{\mathrm{Re}}$. Using the simulation parameters $\sigma_T$ and $\tau$ together with the corresponding values of $T_w$, $T^*$, $\mathrm{Kn}_\infty$ for each configuration, we calculate the model output and compare it to the Nusselt numbers from the numerical simulations. In figure 4, we present the Nusselt number obtained from the model ($\mathrm{Nu}_{\mathrm{model}}$ with $\varepsilon \sim 0.2$, best fit) plotted against the numerical results of Nu. The model provides a remarkable match with the simulation data. The mean error of the estimated Nusselt number for all values of $\sigma_T$ and $\tau$ was less than 3% in the Knudsen range up to 0.06. It is worth mentioning that variations of $T_w$ on the order of 20% would only impact the value of Nu by 2%. This justifies our previous consideration $T_g \sim T_w$ in equation 15.

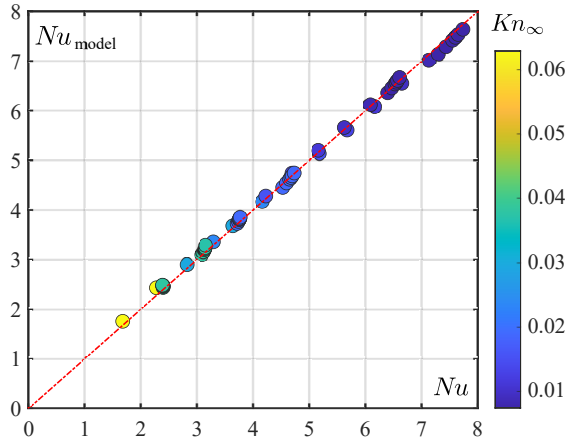

Figure 4: Correlation between the modeled dimensionless heat flux $Nu_{model}$ and the Nusselt number obtained from the numerical simulations Nu (temperature ratio parameter $\varepsilon \sim 0.2$).

Table 1: Thermal accommodation and temperature jump coefficients

| $\sigma_T$ ($\pm 0.02$) | 0.65 | 0.75 | 0.77 | 0.83 |
|---|---|---|---|---|
| $\zeta_T$ ($\pm 0.05$) | 3.07 | 2.46 | 2.36 | 2.08 |
| $T_w$ ($\pm 3\,\mathrm{K}$) | 495 | 439 | 390 | 356 |

A number of implications can be deduced from the model. First, Nu $\propto$ Re asymptotically for large Kn, in accordance to free-molecular flow theory (Baldwin et al., 1960). Secondly, the effect of the temperature loading $\tau$ is counter-balanced by $\sigma_T$: while $Nu_c$ increases with $\tau$, this augmentation is not followed by Nu in slip flow because $\phi$ increases for small $\sigma_T$. This helps elucidating the intriguing observation pointed out by Kovasznay (1950) who found that Nu decreases with $\tau$ for supersonic flows, in contrast to the results in low-speed conditions (Collis and Williams, 1959; Smits et al., 1983). Indeed, it is well known that $\sigma_T$ decreases with the surface temperature (Grilly et al., 1946). According to the model, this would imply a reduction of Nu with increasing overheat ratio. Finally, as Nu is an explicit function of $\sigma_T$, the model provides a method to extract the gas-surface thermal accommodation coefficient from measurements of heat loss and known values of the continuum Nusselt number $Nu_c$.

With the goal of determining the thermal accommodation coefficient using the model, we employ equations 14 and 15 plugging the values of Nu extracted from Laufer and McClellan (1956), and additional measurements of the current campaign with lower wire temperatures. The calculated values of $\sigma_T$ are shown in figure 5. Within the measurement uncertainties, the values of $\sigma_T$ for the air-platinum interface compare very well by applying the current experimental data and the measurements from Laufer and McClellan (1956). They are also in excellent agreement with the trend reported by Grilly et al. (1946), where $\sigma_T$ decreases monotonically for increasing surface temperature up to 600 K.

It is of interest to note that the methodology described above can be employed to determine more general temperature jump coefficients of the form:

$$T_g - T_w = \zeta_T \lambda_e \frac{\partial T}{\partial n}\bigg|_{wall}. \qquad (16)$$

Here, $\zeta_T$ is the temperature jump coefficient, and $\lambda_e$ is the equivalent free path of the gas given by $\lambda_e = (2/\sqrt{\pi})\lambda$ (Sharipov, 2011; Su et al., 2022). From the values of $\sigma_T$ obtained in our experiments and shown in figure 5, by using direct comparison of equations 16, 8 and 9, we calculate the temperature jump coefficients $\zeta_T$ and report them in table 1. The values of $\zeta_T$ obtained in this way are in excellent agreement with the temperature jump coefficients reported by Sharipov (2011) for a unitary tangential momentum accommodation, and indicate an increase of $\zeta_T$ with the surface temperature. This methodology paves the way for future measurements of thermal accommodation or jump coefficients in monoatomic and diatomic gases, and provides a complementary technique to current experimental methods (Yamaguchi et al., 2012; Sharipov, 2011).

## 4. Conclusions

The effect of a temperature jump condition on the heat transfer from a small cylinder was examined in the supersonic slip flow regime. Combined numerical simulations and hot-wire measurements revealed the significant impact of the thermal accommodation coefficient on the heat loss from a cylinder in supersonic flow. Using the results issued from the numerical simulations in the framework of the Smoluchowski temperature jump condition, we derived and verified a simple model that explicitly predicts the Nusselt number of a heated cylinder as a function of the Knudsen number and the thermal accommodation coefficient, additionally accounting for the cylinder overheating parameter. The model puts in evidence the competing effects between thermal accommodation and wall temperature, clarifying the decrease of the Nusselt number with the overheating parameter observed by previous authors in supersonic flow (Kovasznay, 1950; Smits et al., 1983).

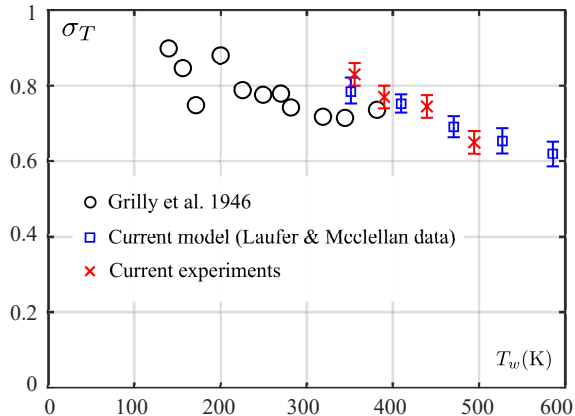

Figure 5: Thermal accommodation coefficient of air-platinum as a function of the wall temperature.

Given the analytical relation obtained for the Nusselt number, $Nu = Nu(Nu_c, Kn, \sigma_T)$, we employed the model reversely to deduce $\sigma_T$ from simultaneous values of Nu, $Nu_c$ and Kn, giving rise to a novel methodology to determine $\sigma_T$ based on classical hot-wire measurements of Nu and known values of the continuum Nusselt number $Nu_c$. The values of air-platinum $\sigma_T$ reported here correspond to an unprecedented explored temperature range, and may provide useful input for the calculation of temperature jump coefficients and validation of *ab initio* theoretical models, in particular for multi-component gas mixtures (Sharipov and Kalempa, 2005; Sharipov, 2024). In future work, this methodology will be extended to subsonic flow conditions due to their inherent simpler experimental implementation, and other gas-surfaces couples could be investigated in the same framework. Beyond the fact that the accommodation coefficient is as important as the classical transport properties in rarefied flows (Sharipov, 2011), they may be relevant, for instance, in the modelling of state-of-the-art micro and nanoscale sensing devices operating in the slip flow regime (Bailey et al., 2010; Roseman and Argrow, 2021; Le-The et al., 2021; Brunier-Coulin et al., 2023).

## 5. Acknowledgements

The authors are thankful to Alejandro Gomez Mesa for the compilation of the heat transfer data. D.B is grateful for the financial support from the *Association Française de Mécanique* (AFM), and for enlightening discussions with Prof. Felix Sharipov on temperature jump coefficients.